\documentclass[prd,amsmath,amssymb,showpacs,superscriptaddress,nofootinbib,twocolumn]{revtex4-1}

\usepackage{graphicx}
\usepackage{epsfig,graphics,subfigure,psfrag,amsmath,amssymb}
\usepackage{lineno}
\usepackage{dcolumn}
\usepackage{bm}
\usepackage{overpic}
\usepackage{multirow}
\usepackage[english]{babel}
\addto{\captionsenglish}{%

}

\newcommand{\jpsi}{J/\psi}
\newcommand{\piz}{\pi^{0}}
\newcommand{\pip}{\pi^{+}}
\newcommand{\pim}{\pi^{-}}
\newcommand{\Kp}{K^{+}}
\newcommand{\Km}{K^{-}}
\newcommand{\BR}{\mathcal{B}}
\newcommand{\pp}{p\bar{p}}
\newcommand{\BB}{B\bar{B}}
\newcommand{\ep}{e^{+}}
\newcommand{\en}{e^{-}}
\newcommand{\psipp}{\psi(3770)}
\newcommand{\BRp}{(2.4\pm0.8\pm0.3)\times10^{-5}}
\newcommand{\BRlam}{(1.7\pm0.6\pm0.1)\times10^{-5}}
\newcommand{\BRsigp}{(4.5\pm0.9\pm0.1)\times10^{-5}}
\newcommand{\BRsigz}{(4.5\pm0.9\pm0.1)\times10^{-5}}
\newcommand{\BRxim}{(2.0\pm0.7\pm0.1)\times10^{-5}}
\newcommand{\BRxiz}{(2.0\pm0.7\pm0.1)\times10^{-5}}
\begin{document}
%\widetext
\normalsize
\parskip=5pt plus 1pt minus 1pt

\title{\boldmath Study of $\psi(3770)$ decaying to Baryon anti-Baryon Pairs}
\author{ Li-Gang Xia\\ \it Department of Physics, Tsinghua University, Beijing 100084, People's Republic of China}
%\input{authorlist}

%\input author_list.tex       % D0 authors (remove the first 3 lines
                             % of this file prior to submission, they
                             % contain a time stamp for the authorlist)
                             % (includes institutions and visitors)
%\date{\today}

\begin{abstract}
To study the decays of $\psi(3770)$ going to baryon anti-baryon pairs ($\BB$), all available experiments of measuring the cross sections of $e^+e^- \to \BB$ at center-of-mass energy ranging from 3.0~GeV to 3.9~GeV are combined. To relate the baryon octets, a model based on the SU(3) flavor symmetry is used and the SU(3) breaking effects are also considered. Assuming the elctric and magnetic form factors are equal ($|G_E|=|G_M|$), a global fit including the interference between the QED process and the resonant process is performed. The branching fraction of $\psipp \to \BB$ is determined to be $\BRp$, $\BRlam$, $\BRsigp$, $\BRsigz$, $\BRxim$, and $\BRxiz$ for $B=p, \Lambda, \Sigma^+, \Sigma^0, \Xi^-$ and $\Xi^0$, respectively, where the first uncertainty is from the global fit and the second uncertainty is the systematic uncertainty due to the assumption $|G_E|=|G_M|$. They are at least one order of magnitude larger than a simple scaling of the branching fraction of $\jpsi/\psi(3686)\to\BB$.
\end{abstract}

\maketitle

%\newpage
%\tableofcontents
%\newpage

%\setpagewiselinenumbers
%\modulolinenumbers[5]
%\linenumbers
\section{Introduction}
The $\psi(3770)$ is the lowest lying $1^{--}$ charmonium state above the charmed meson pair threshold. It decays dominantly into $D^0\bar{D}^0/D^+D^-$ while the decays to the light hadron (LH) final states are OZI-suppressed.
It is still unclear about the nature of $\psi(3770)$. If it is a pure $c\bar{c}$ bound state, the branching fraction of $\psipp$ into non-$D\bar{D}$ decays ranges from less than 1\% from the potential models~\cite{potential1,potential2} to about 5\% from the non-relativistic QCD calculations~\cite{NRQCD1,NRQCD2}. If $\psi(3770)$ has a four-quark admixture, the total non-$D\bar{D}$ branching fraction could be up to 10\%~\cite{fourq}.

Experimentally, the BES collaboration reported a large non-$D\bar{D}$ branching fractions of $(14.5\pm1.7\pm5.8)\%$~\cite{nonDDbes1,nonDDbes2,nonDDbes3} neglecting the interference between the $\psi(3770)$ resonant amplitude and the QED continuum amplitude. Only considering the interference between the one-photon amplitude of the $\psi(3770)$ resonance and the QED continuum amplitude, the CLEO collaboration found this branching fraction to be $(-3.3\pm1.4_{-4.8}^{+6.6})\%$~\cite{nonDDcleo}. To clarify the disagreement, many exclusive non-$D\bar{D}$ decays with the light hadron final state have been searched for using two methods~\cite{cleoBB,cleoLH,besLH,bes3BB,bes3psipp,bes3psipp}. One method is to compare the cross section at the center-of-mass (c.m.) energy ($\sqrt{s}$) close to the $\psi(3770)$ nominal mass and that far from any charmonium resonance (for example, the two energies are $3.773$~GeV and $3.671$~GeV for the CLEO collaboration). Only for the final state $\phi \eta$, there is a significantly excessive cross section at $\sqrt{s}=3.773$~GeV~\cite{cleoLH}. The other method, allowing to consider the complicated interference effect, is to perform a scan around the $\psi(3770)$ resonance. Using this methd, the BESIII collaboration reports that the line shape of the cross section shows a deficit in the vicinity of the $\psi(3770)$ for the final states $\pp$ and $\pp\piz$~\cite{bes3psipp,bes3pppiz}. Furthermore, there is a two-solution ambiguity for the branching fraction of $\psi(3770) \to \pp/\pp\piz$, which cannot be solved from the scan experiment. Recently, an evidence of $\psi(3770) \to \Kp\Km$ was also found by studying the cross section of $\ep\en \to \Kp\Km$ above 2.6~GeV~\cite{psippKK}.

We focus on the decays of $\psi(3770)$ going to baryon anti-baryon pairs ($\BB$). Here $B=p, \Lambda,\Sigma^+, \Sigma^0, \Xi^-$ and $\Xi^0$. All available experiments measuring the cross section of $\ep\en\to\BB$ at the c.m. energy from 3~GeV to 3.9~GeV are combined. In Sec. II, we will present the born cross section formulas of $e^+e^- \to \BB$ and introduce the model to relate all the baryon octet states. In Sec. III, we will review the available experiments and describe the fit strategy. The results will be shown and discussed in Sec. IV. A short summary will be given in Sec. V.

\section{Cross section formulae of $e^+e^- \to \BB$ and description of the model}
The Born cross section of the QED process $\ep\en \to \gamma^* \to \BB$ at the center-of-mass energy $\sqrt{s}$ can be written as
\begin{equation}~\label{eq:cross_qed}
  \sigma_{QED}(s) = \frac{4\pi\alpha^2\beta_B}{3s}\left(|G_M^B(s)|^2 + \frac{2M_B^2}{s}|G_E^B(s)|^2\right) \: ,
\end{equation}
where $M_B$ is the nominal baryon mass, $\beta_B \equiv \sqrt{1-4M_B^2/s}$, $\alpha$ is the fine-structure constant, and $G_M^B$ and $G_E^B$ are the magnetic and electric form factors~\cite{rmpFF}, respectively. 

The resonance production cross section of $\ep\en\to\psipp\to\BB$ is written as
\begin{equation}
  \sigma_{RES} = \frac{12\pi\Gamma_e\Gamma_B}{(s-M_0)^2+M_0^2\Gamma_0^2} \: ,
\end{equation}
where $M_0=3773.15$~MeV/$c^2$ and $\Gamma_0=27.2$ MeV~\cite{PDG} are the nominal mass and total width of $\psi(3770)$, $\Gamma_e$ ($\Gamma_B$) is the partial width of $\psipp \to \ep\en$ ($\BB$). $\Gamma_B$ can be written as
\begin{equation}\label{eq:GammaB}
  \Gamma_B = \frac{M_0}{12\pi}\sqrt{1-\frac{4M_B^2}{M_0^2}}\left(|F_M^B|^2 + \frac{2M_B^2}{M_0^2}|F_E^B|^2 \right) \: ,
\end{equation}
where $|F_M^B|$ and $|F_E^B|$ are the form factors. 

The form factor ratio $|G_E/G_M|$ is 1 at the baryon pair threshold, but may have small deviations above the threshold. The predicted behavior is model-dependent (see for example Ref.~\cite{dubnicka, kang}). Experimentally, the form factor ratio is measured to be consistent with 1 within the uncertainties for the proton~\cite{babarLA,bes3pp} in the region $2.2<\sqrt{s}<3.1$ GeV and for the baryon $\Lambda$~\cite{babarBB} in the mass region from the threshold to 2.8 GeV. However, the measurement of the neutron form factor from the threshold up to 2.44 GeV~\cite{fenice} indicates $|G_E|=0$. Throughout this paper, we assume that $|G_{E}^B|=|G_{M}^B|\equiv |G^B|$ and $|F_E^B|=|F_M^B|\equiv |F^B|$ for all $\BB$ final states. The effect of this assumption will be considered. The  nucleon electromagnetic form factors in the timelike region have been extensively reviewed in Ref.~\cite{denig}. Here, the form factors $G^B$ and $F^B$ take the following forms
\begin{equation}
  G^B(s) = \frac{C^B}{s^2\ln^2(s/\Lambda^2)} \: ,
\end{equation}
from a calculation in Ref.~\cite{Gs} and
\begin{equation}\label{eq:FB}
  F^B = \sqrt{\frac{12\pi\Gamma_e}{M_0}}G^B(M_0^2)e^{i\phi'} + A^Be^{i\phi} \: .
\end{equation}
Here $\Lambda=0.3$~GeV is the QCD scale parameter, $C^B$ and $A^B$ are the free parameters. In Eq.~\ref{eq:FB}, the first term represents the electromagnetic interaction amplitude of the $\psi(3770)$ and the second term represents the OZI-suppressed strong decay amplitude of the $\psi(3770)$. Two phase angles $\phi'$ and $\phi$ are introduced relative to the QED process. $\phi'$ represents the phase difference between the electromagnetic amplitude of the $\psi(3770)$ resonance and the QED continuum amplitude. In many analyses (for example Ref.~\cite{bes3psipp,kzhu}), this phase difference is assumed to be 0, namely, $\phi'=0$. We will find that the effect of the nonzero $\phi'$ is also negligible in the case of $\psi(3770) \to \BB$. 

Therefore, the total cross section considering the interference between the processes $\ep\en \to \gamma^* \to \BB$ and $\ep\en \to \psipp\to\BB$ is constructed as
\begin{eqnarray}
  \sigma_{TOT} = &&\frac{4\pi\alpha^2\beta_B}{3s}\left|\sqrt{1+\frac{2M_B^2}{s}}\frac{C^B}{s^2ln^2(s/\Lambda^2)}\right. \nonumber \\ &&+\left.3\sqrt{s}\frac{\sqrt{\Gamma_e/\alpha^2}\sqrt{\Gamma_B/\beta_B}}{s-M_0^2 + iM_0\Gamma_0}\frac{F^B}{|F^B|}\right|^2 \: .
\end{eqnarray}

To relate the form factors for all baryon octets, the SU(3) flavor symmetry is imposed. We also consider the SU(3) breaking effect due to the electromagnetic interaction and the quark mass difference of $m_s-m_{u/d}$. For convenience, we introduce the matrix notations. The SU(3) octet baryons and anti-baryons are described by the matrices $\bold{B}$ and $\bold{\bar{B}}$ respectively.
\begin{equation}
  \bold{B}=\begin{pmatrix} \Sigma^0/\sqrt{2}+\Lambda/\sqrt{6}& \Sigma^+ & p \\
  \Sigma^- & -\Sigma^0/\sqrt{2}+\Lambda/\sqrt{6} & n \\
  \Xi^- & \Xi^0 & -2\Lambda/\sqrt{6} \end{pmatrix}
\end{equation}
\begin{equation}
  \bold{\bar{B}}=\begin{pmatrix} \bar{\Sigma}^0/\sqrt{2}+\bar{\Lambda}/\sqrt{6}& \bar{\Sigma}^- & \bar{\Xi}^- \\
  \bar{\Sigma}^+ & -\bar{\Sigma}^0/\sqrt{2}+\bar{\Lambda}/\sqrt{6} & \bar{\Xi}^0 \\
  \bar{p} & \bar{n} & -2\bar{\Lambda}/\sqrt{6} \end{pmatrix}
\end{equation}
 The SU(3) invariant effective lagrangian for the decay $\psi(3770) \to \BB$ can be written as
 \begin{eqnarray}\label{eq:Leff}
   \mathcal{L}_{eff}=&&gTr(\bold{B}\bar{\bold{B}})  \nonumber\\
   &&+ dTr(\{\bold{B},\bar{\bold{B}}\}\bold{S}_e)+
   fTr([\bold{B},\bar{\bold{B}}]\bold{S}_e) \nonumber \\
   &&+ d'Tr(\{\bold{B},\bar{\bold{B}}\}\bold{S}_m)+
   f'Tr([\bold{B},\bar{\bold{B}}]\bold{S}_m) \:,
 \end{eqnarray}
 where $g,d,f,d',f'$ are the coupling constants, ``$Tr$'' represents the trace of a matrix, ``[a,b]'' and ``\{a,b\}'' denote the commutator and the anticommutator of the two elements $a$ and $b$ respectively, and the matrices $\bold{S}_e$ and $\bold{S}_m$ are defined as
 \begin{equation}
   \bold{S}_e = \begin{pmatrix}
2 & & \\
& -1 & \\
& & -1
   \end{pmatrix}
   \: ,
   \bold{S}_m = \begin{pmatrix}
     1 & & \\
     & 1 & \\
     & & -2
   \end{pmatrix} \:.
 \end{equation}
 In the right-hand side of Eq.~\ref{eq:Leff}, the first line represents the OZI-suppressed strong amplitude, the second line represents the one-photon electromagnetic amplitude, and the third line represents the SU(3)-breaking contribution due to the quark mass difference (more details about the effective lagrangian can be found in Ref.~\cite{BBPars1,BBPars2,kzhu}).

 From Eq.~\ref{eq:Leff}, we can derive the following relations for the form factors $G^B$ and $F^B$ (or equivalently $C^B$ and $A^B$).
\begin{eqnarray}
  && C^p = C_1+C_2 \nonumber \\
  && C^\Lambda = -C_1 \nonumber \\
  && C^{\Sigma^+}= C_1+C_2 \nonumber \\
  && C^{\Sigma^0}= C_1 \nonumber \\
  && C^{\Xi^-} = C_1-C_2 \nonumber\\
  && C^{\Xi^0} = -2C_1 \nonumber
\end{eqnarray}
and
\begin{eqnarray}
   &&A^p = A_0 - A_1 + A_2 \nonumber \\
   &&A^{\Lambda} = A_0 - 2A_1 \nonumber \\
   &&A^{\Sigma^+} = A_0 + 2A_1 \nonumber \\
   &&A^{\Sigma^0} = A_0 + 2A_1 \nonumber \\
   &&A^{\Xi^-} = A_0 - A_1 - A_2 \nonumber \\
   &&A^{\Xi^0} = A_0 - A_1 - A_2 \nonumber
\end{eqnarray}
Here the free parameters $C_1, C_2, A_0, A_1, A_2$ are real numbers in practice.

\section{Experimental review and the fitting strategy}
\subsection{Experimental review}
We starts with the reaction $\ep\en \to \pp$ for which the most data sets have been accumulated. The BESIII collaboration has performed a scan from 3.65~GeV to 3.90~GeV and a deficit is found in the vicinity of the $\psi(3770)$~\cite{bes3psipp}. Considering the interference between the QED process and the $\psi(3770)$ resonant production, two solutions are found for the partial width of $\psi(3770) \to \pp$ with equal goodness of fit. But Ref.~\cite{bes3psipp} has not reported the statistical significance of the solutions. To solve this two-solution ambiguity, more experimental information is needed. The results from the studies of the proton form factors from the CLEO~\cite{cleo1,cleo2}, the BES/BESIII~\cite{bespp,bes3pp} and the BABAR~\cite{babarLA,babarSA} collaborations can be used. The former two collaborations measure the cross section of $\ep\en \to \pp$ using Eq.~\ref{eq:bes}.
\begin{equation}\label{eq:bes}
  \sigma_i^{Born} = \frac{N_i^{obs}}{L_i\epsilon_i(1+\delta_i)} \: ,
\end{equation}
where $i$ denotes the energy point, $N_i^{obs}$ is the observed number of signal events, $L_i$ is the luminosity, $\epsilon_i$ is the efficiency, and $(1+\delta_i)$ is the radiative correction factor~\cite{rc1,rc2,rc3}.
The BABAR collaboration utilizes the initial state radiation (ISR) technique~\cite{rmpISR}. The process is $\ep\en \to \gamma \pp$, where the photon can be required to be detected~\cite{babarLA} or undetected~\cite{babarSA}. The cross section of $\ep\en \to \pp$ at the c.m. energy of the $\pp$ invariant mass $M_{\pp}$ can be extracted according to Eq.~\ref{eq:babar}.
\begin{equation}\label{eq:babar}
\sigma(M_{\pp}) = \frac{(dN/dM_{\pp})_{corr}}{(dL/dM_{\pp})\epsilon(M_{\pp})R} \: ,
\end{equation}
where $(dN/dM_{\pp})_{corr}$ is the mass spectrum corrected for the mass resolution effect, $dL/dM_{\pp}$ is the ISR differential luminosity~\cite{rmpISR}, $\epsilon(M_{\pp})$ is the detection efficiency, and $R$ is the radiative correction factor.

For the final states $\Lambda\bar{\Lambda}, \Sigma^+\bar{\Sigma}^-$, $\Sigma^0\bar{\Sigma}^0$, $\Xi^-\bar{\Xi}^+$ and $\Xi^0\bar{\Xi}^0$, the CLEO and BES/BESIII collaborations~\cite{cleoBB,besLH,bes3BB} have measured the cross sections at the peak of the $\psi(3770)$ resonance. Neglecting the interference effect with the QED process $\ep\en \to \gamma^* \to \BB$, there is no significant excess compared to the cross section at an energy point far from any charmonium resonance. The BABAR collaboration also studied $\ep\en \to \Lambda\bar{\Lambda}/\Sigma^0\bar{\Sigma}^0/\Lambda\bar{\Sigma}^0$ using the ISR technique and provided the upper limit of the cross section at the 90\% confidence level (CL), which will be used as a cross-check for our results. All the data sets used in the following fit are summarized in Table~\ref{tab:review_pp} and Table~\ref{tab:review_BB}. The denotations for the $\pp$ final state in the first column of Table~\ref{tab:review_pp} will be used consistently throughout this paper.

\begin{table}[htbp]
\center{}
\caption{\label{tab:review_pp} Review of the experiments of measuring the cross section of $\ep\en \to \pp$ from 3~GeV to 3.9~GeV.}
\begin{tabular}{l l l}
\hline\hline
Denotation & Process & $M_{\pp}$ (GeV) \\
\hline
BES~\cite{bespp} & $\ep\en \to \pp$ & 3.0, 3.07 \\
\multirow{2}{*}{ESIII~\cite{bes3pp}} & \multirow{2}{*}{$\ep\en\to\pp$} &  3.05, 3.06, 3.08, 3.40, 3.50\\
& &3.5507, 3.6002, 3.671 \\
\multirow{4}{*}{$\psi''$ scan~\cite{bes3psipp}} & \multirow{4}{*}{$\ep\en \to \pp$} & 3.65, 3.748, 3.752, 3755\\
 & & 3760, 3.766, 3.772, 3.773 \\
 & & 3.778, 3.784, 3.791, 3.798 \\
 & & 3.805, 3.810, 3.819, 3.900\\
BaBar (LA)~\cite{babarLA}& $\ep\en\to\gamma\pp$ & 3.0-4.0 \\
BaBar (SA)~\cite{babarSA}& $\ep\en\to\gamma\pp$ & 3.0-4.0 \\
CLEO~\cite{cleo1,cleo2} & $\ep\en\to\pp$ & 3.671, 3.772 \\
\hline\hline
\end{tabular}
\end{table}

\begin{table}[htbp]
\center{}
\caption{\label{tab:review_BB} Review of the experiments of measuring the cross section of $\ep\en \to \BB$ from 3~GeV to 3.9~GeV.}
\begin{tabular}{l l l}
\hline\hline
Denotation & Process & $M_{\BB}$ (GeV) \\
\hline
CLEO~\cite{cleoBB} & $\ep\en\to\Lambda\bar{\Lambda}\to \pp\pip\pim$ & 3.671, 3.772 \\
\hline
\multirow{4}{*}{BESIII~\cite{bes3BB}} & $\ep\en \to \Sigma^+\bar{\Sigma}^-\to \pp4\gamma$ & \multirow{4}{*}{3.65, 3.773}\\
 & $\ep\en \to \Sigma^0\bar{\Sigma}^0 \to \pp\pip\pim2\gamma$ & \\
 & $\ep\en \to \Xi^-\bar{\Xi}^+\to \pp 2\pip2\pim$ & \\
 & $\ep\en \to \Xi^0\bar{\Xi}^0 \to \pp\pip\pim4\gamma$ &\\
\hline
\multirow{3}{*}{BaBar~\cite{babarBB}} & $\ep\en\to\gamma(\Lambda\bar{\Lambda}) \to \gamma(\pp\pip\pim)$ & 3.2-3.6\\
& $\ep\en \to \gamma (\Sigma^0\bar{\Sigma}^0) \to \gamma(\pp\pip\pim2\gamma)$ & 3.2-3.6 \\
& $\ep\en \to \gamma (\Lambda\bar{\Sigma}^0) \to \gamma(\pp\pip\pim\gamma)$ & 2.9-3.8 \\
\hline\hline
\end{tabular}
\end{table}

\subsection{The fitting strategy}
To combine the results from various experiments, we should consider the statistical uncertainties and the systematical uncertainties correctly. For the number of signal events $N^{obs}$, it is either obtained by simply neglecting the background and counting the number of events or extracted by subtracting the background events from the total number of events. Either way leads to a systematical uncertainty. At all energy points in an experiement, the luminosities are measured using the same method, the signal events are selected using the same set of conditions, and the radiative correction factors are obtained in the same way. Thus the systematical uncertainties related to them are independent upon the energy point and will be considered by introducing a free normalization factor for each experiment.

We starts with the case of proton. A $\chi^2$ is constructed in Eq.~\ref{eq:chi2} for each experiment except for the ``$\psi''$ scan'' experiement~\cite{bes3psipp}.
\begin{equation}\label{eq:chi2}
  \chi_{\alpha}^2(p) = \frac{(1-f_\alpha)^2}{\xi_{ind.}^2} + \sum_{i_\alpha} \frac{(N_{i_\alpha}^{obs}-f_{\alpha}\lambda_{i_\alpha})^2}{(\Delta N_{i_\alpha}^{obs})_{tot.}^2 + (\Delta \epsilon/\epsilon f_\alpha \lambda_{i_\alpha})^2} \: ,
\end{equation}
where $\alpha$ denotes the experiment, $i_\alpha$ denotes the $i$-th energy point for the experiment $\alpha$, $N^{obs}$ is the observed number of signal events, $\lambda$ is the expected number of signal events and defined as $\lambda \equiv \sigma L\epsilon(1+\delta)$ or $\sigma L \epsilon R$ as indicated in Eq.~\ref{eq:bes} and Eq.~\ref{eq:babar}. $(\Delta N^{obs})_{tot.}$ is the quadratic sum of the statistical uncertainty of $N^{obs}$ and the systematical uncertainty due to the background subtraction or neglecting the background events. $\Delta \epsilon$ is the statistical uncertainty of the efficiency $\epsilon$ determined from a limited MC sample. $\xi_{ind.}^2$ is the quadratic sum of the systematical uncertainties which are independent upon the energy point. It includes the systematical uncertainties due to the consistent selection criteria at all energy points, the trigger efficiency, the reconstruction efficiency of charged tracks, the efficiency corrections as used in the BABAR measurements~\cite{babarLA,babarSA}, the measurement of the luminosities, and the radiative correction factors. To consider these systematical uncertainties independent upon the energy point, the free normalization factor $f_\alpha$ is introduced for each experiment. Here two things should be noted. One is that we do not consider the correlation of various selection conditions. The other is that we assume the form factors satisfy $|G_E|=|G_M|$ and thus we do not consider the efficiency uncertainty due to this assumption (typically, the efficiency with $|G_E|=|G_M|$ is $5\%-10\%$ different from that with $|G_E|=0$~\cite{babarSA}).

For the ``$\psi''$ scan'' experiment in which $N^{obs}$ is found to be 0 at some energy points and the background contamination is only 0.6\%, it is better to construct the likelihood function assuming that the number of signal events at each energy point abides by the poisson distribution as shown in Eq.~\ref{eq:poisson}.
\begin{equation}\label{eq:poisson}
\chi_{P}^2(p) \equiv \frac{(1-f)^2}{\xi_{ind.}^2}-2\sum_i \ln P(N_i^{obs}|f\lambda_i) - \ln P(N_i^{obs}|N_i^{obs}) \: ,
\end{equation}
where $P(N|\nu)$ is the probability of observing $N$ events with the expectation value $\nu$ in the poisson distribution, namely, $P(N|\nu)\equiv \nu^Ne^{-\nu}/N!$, and $f$ is the free normalization factor.

For other baryon octets, the cross section of $\ep\en \to \Lambda\bar{\Lambda}$ at the peak of the $\psi(3770)$ reported by the CLEO collaboration~\cite{cleoBB} and that of $\ep\en \to \Sigma^+\bar{\Sigma}^-/\Sigma^0\bar{\Sigma}^0/\Xi^-\bar{\Xi}^+/\Xi^0\bar{\Xi}^0$ reported by the BESIII collaboration~\cite{bes3BB} are used. As shown in the second column of Table~\ref{tab:review_BB}, these processes share some final particles such as protons, pions and photons. The related systematical uncertainties due to the reconstruction of proton and pion tracks, the second vertex fit, the particle identification, and the detection of the photons are shared. However, Ref.~\cite{bes3BB} did not report the individual systematical uncertainties. It is impossible to treat them correctly. Fortunately, the limited knowledge of the angular distribution contributes the dominant systematical uncertainty of $9.2\%-10.9\%$, which depends upon the baryon pairs and should be considered individually. The $\chi^2$ is then constructed as follows.
\begin{equation}\label{eq:chi2BB}
\chi^2(B) = \sum_{B=\Lambda,\cdots, \Xi^0} \frac{(N_{B}^{obs} - \lambda_B)^2}{(\Delta N_B^{obs})_{tot.}^2 + (\Delta \lambda_B)_{tot.}^2} \: ,
\end{equation}
where $B$ denotes the baryon, $\lambda$ is the expected number of signal events and defined as $\lambda = \sigma L \epsilon (1+\delta) \times \BR_f$ with $\BR_f$ being the product of the branching fractions of the intermediate-state decays, and $(\Delta \lambda)_{tot.}$ is the total uncertainty of the expected number of signal events.

To combine all experiments, the full optimization quantity is defined as $\chi_{full}^2 \equiv \sum_\alpha \chi_\alpha^2(p) + \chi_{P}^2(p) + \chi^2(B)$. $A_{0,1,2}$, $C_{1,2}$, $f_{\alpha}$, $\phi'$ and $\phi$ are the free parameters.

\section{Fit results and discussions}
\subsection{Fit to the cross section of $\ep\en \to \pp$}
\begin{figure}[htbp]
  \includegraphics[width = 0.45\textwidth]{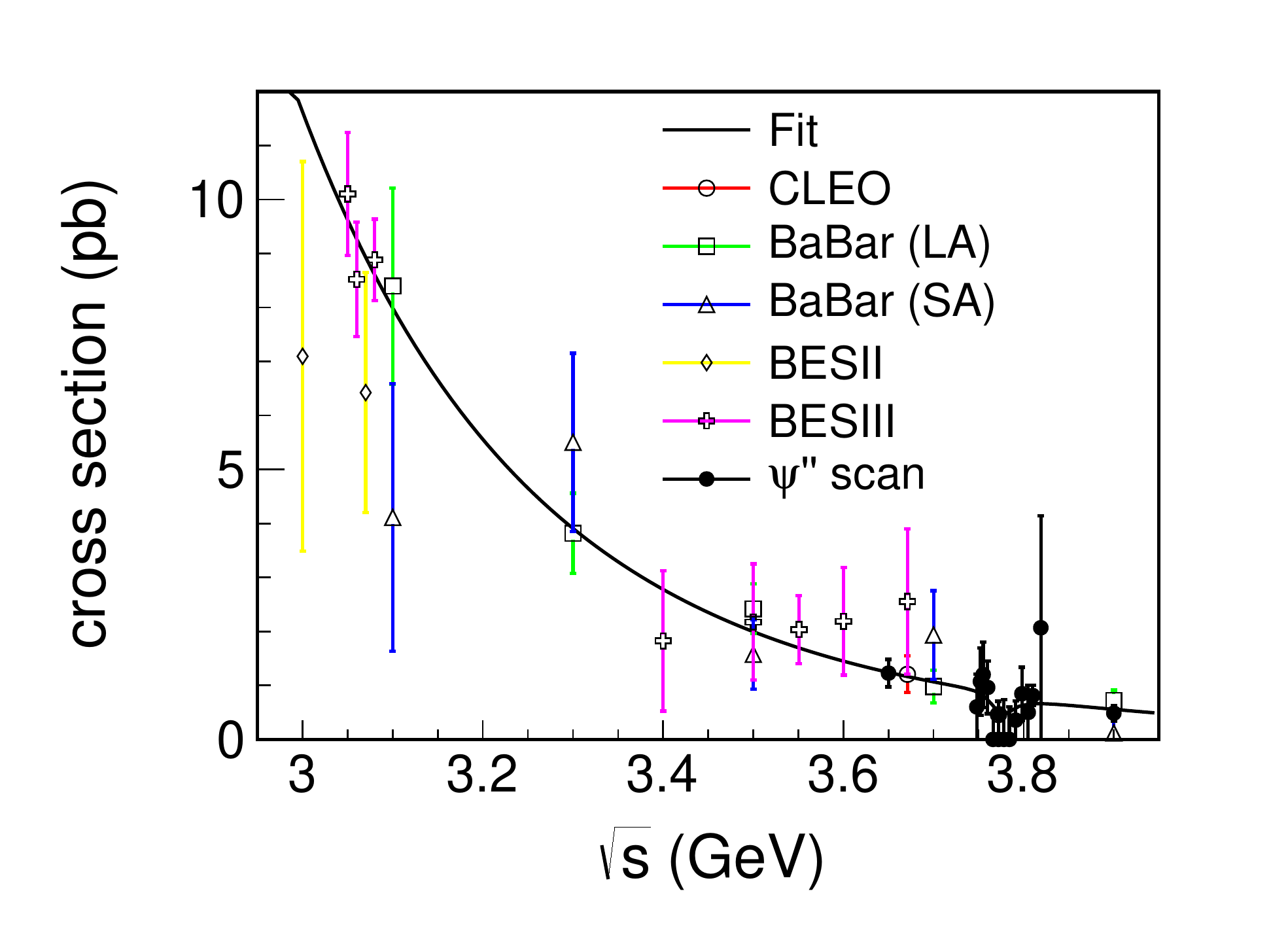}
   \put(-160, 130){\textbf{(a)}}\\
  \includegraphics[width = 0.45\textwidth]{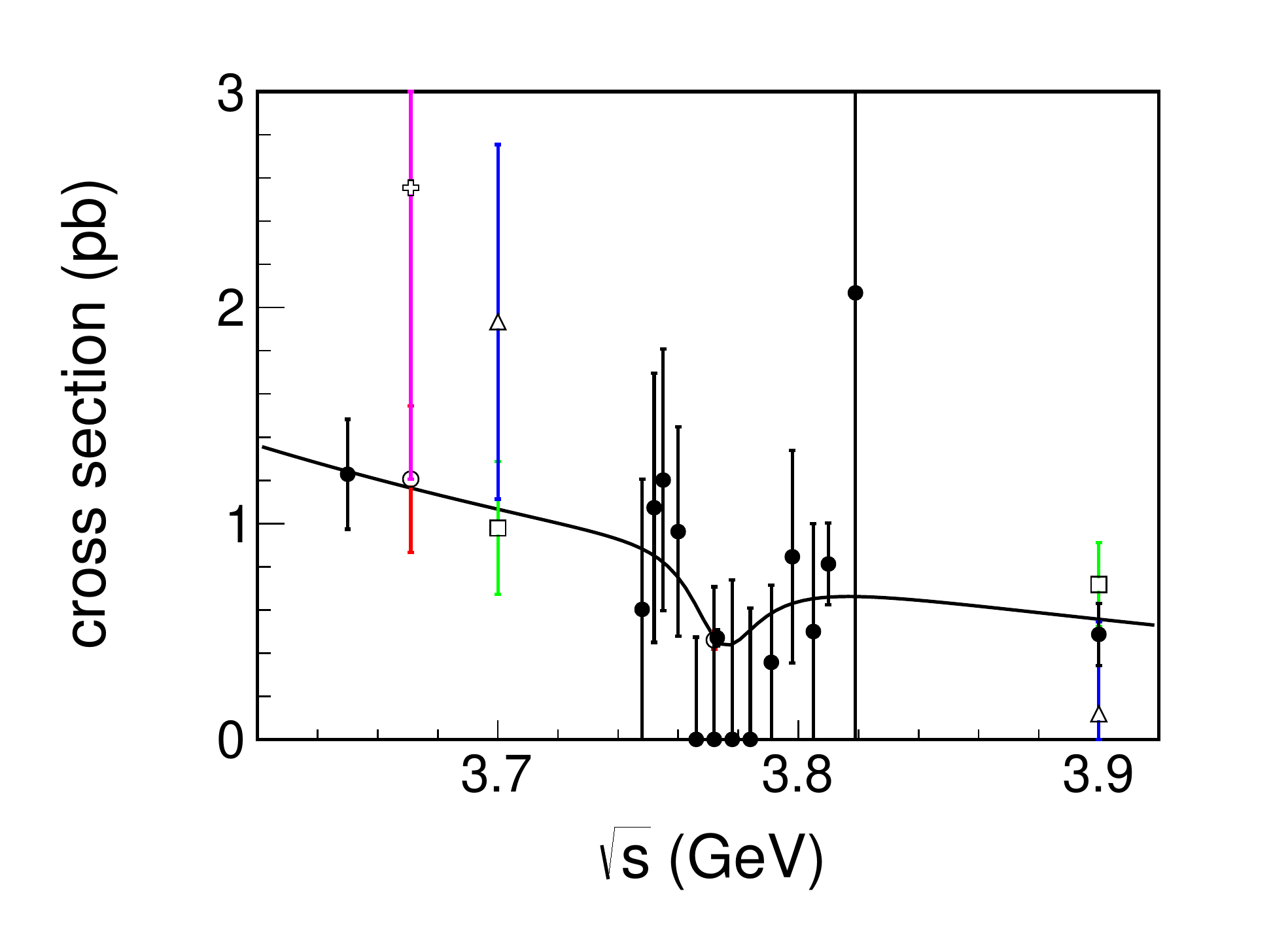}
   \put(-160, 130){\textbf{(b)}}\\
  \caption{The cross section of $\ep\en \to \pp$ and the fit. (b) is the zoom of (a) around the $\psi(3770)$ resonance. Dots with error bar are the data. The black curve is the theoretic prediction from the fit.}\label{fig:fitpp}
\end{figure}

At first, we try the fit in the case of $\pp$. The free parameters are $f_\alpha$, $A^p$, $C^p$ and $\phi$. Here, $\phi'$ is fixed to be 0 for two reasons. One is that we can directly compare the result and that from Ref.~\cite{bes3psipp}. The other is that floating $\phi'$ leads to negligible difference. Two solutions are found with the same goodness of fit $\chi^2/ndof = 25.9/29$, where $ndof$ is the number of degree of freedom. The branching fraction of $\psi(3770) \to \pp$ is found to be either $(6.8_{-2.2}^{+7.1})\times10^{-6}$ or $(2.5\pm0.1)\times10^{-4}$. If the process of $\psi(3770) \to \pp$ is not included, the fit gives $\chi^2/ndof = 91.5/31$, which means that the statistical significance of both solutions is larger than 5 standard deviations. Our results, summarized in Table~\ref{tab:fitpp1} and Table~\ref{tab:fitpp2}, are consistent with those in Ref.~\cite{bes3psipp}. But Ref.~\cite{bes3psipp} does not report the statistical significance of the solutions. Figure~\ref{fig:fitpp} shows the cross sections of $\ep\en\to\pp$ from various experiments and the fit. There is an obvious deficit in the vicinity of the $\psi(3770)$.

\begin{table}[htbp]
\center{}
\caption{\label{tab:fitpp1} Two solutions from fitting to the cross sections of $\ep\en \to \pp$}
\begin{tabular}{l c c c}
\hline\hline
$\BR(\psipp \to \pp)$ & $A^p$ &  $C^p$ (GeV$^4$) & $\phi$\\
\hline
$(6.8_{-2.2}^{+7.1})\times10^{-6}$ & $1.4_{-0.3}^{+0.6}$ & $61.7\pm1.5$ & $(-109^\circ\pm28.3^\circ)$  \\
$(2.5\pm0.1)\times10^{-4}$ & $8.4\pm0.2$ &$61.7\pm1.5$ & $(-94.4^\circ\pm5.3^\circ)$ \\
\hline\hline
\end{tabular}
\end{table}

\begin{table}[htbp]
\center{}
\caption{\label{tab:fitpp2} Normalization factors from fitting to the cross sections of $\ep\en \to \pp$}
\begin{tabular}{ll}
\hline\hline
Normalization factor & Value\\
\hline
$f_{CLEO}$ & $1.00\pm0.02$ \\
$f_{Babar SA}$ & $1.01 \pm 0.04$ \\
$f_{Babar LA}$ & $0.99 \pm 0.05$ \\
$f_{BES}$ & $0.99 \pm 0.05$ \\
$f_{BESIII}$& $1.00 \pm 0.03$\\
$f_{\psi'' scan}$ & $1.02\pm 0.05$ \\
\hline\hline
\end{tabular}
\end{table}
\subsection{Fit to the cross sections of $\ep\en \to \BB$}
\begin{figure*}[htbp]
  \includegraphics[width = 0.45\textwidth]{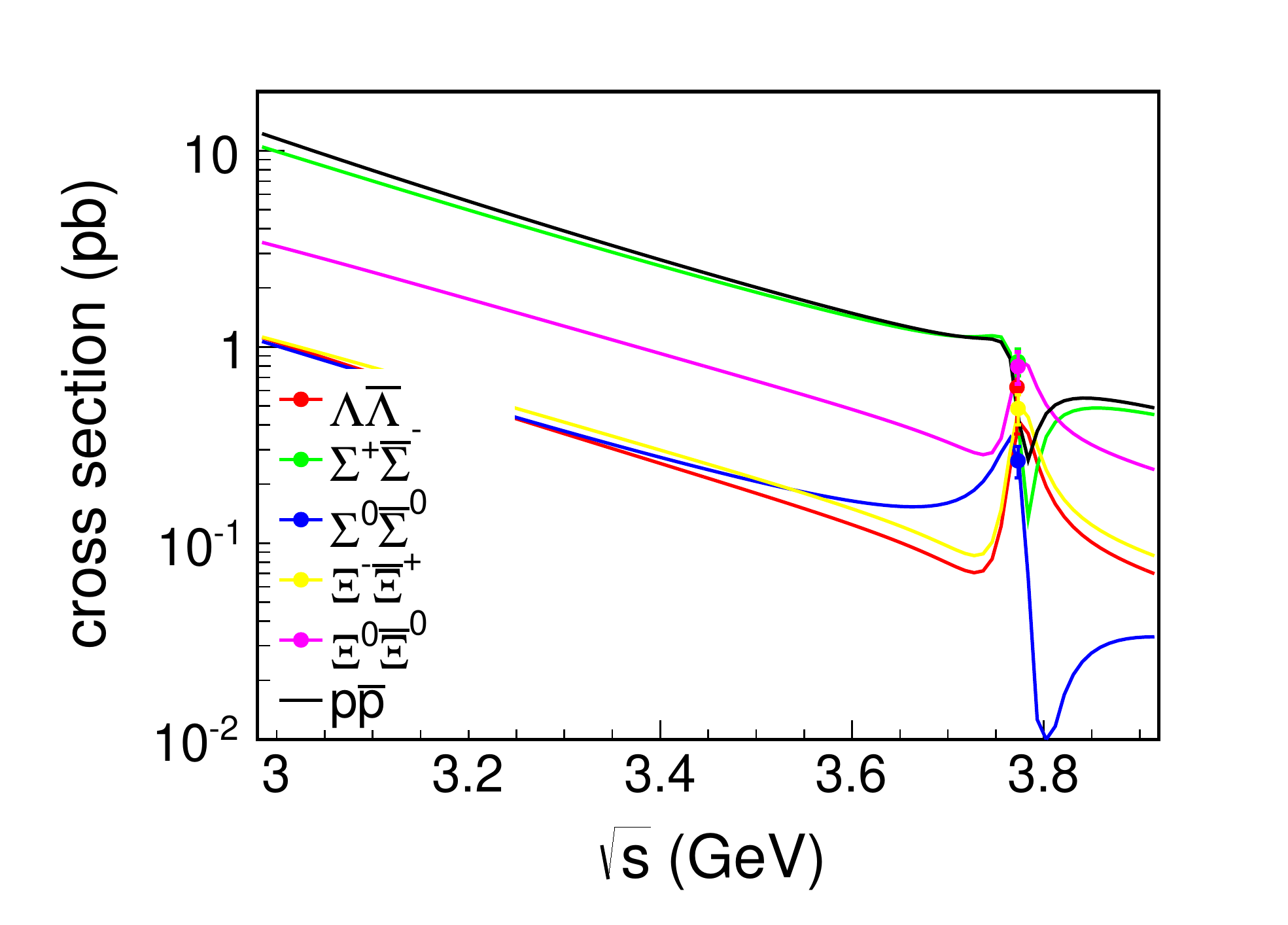}
   \put(-160, 130){\textbf{(a)}}
  \includegraphics[width = 0.45\textwidth]{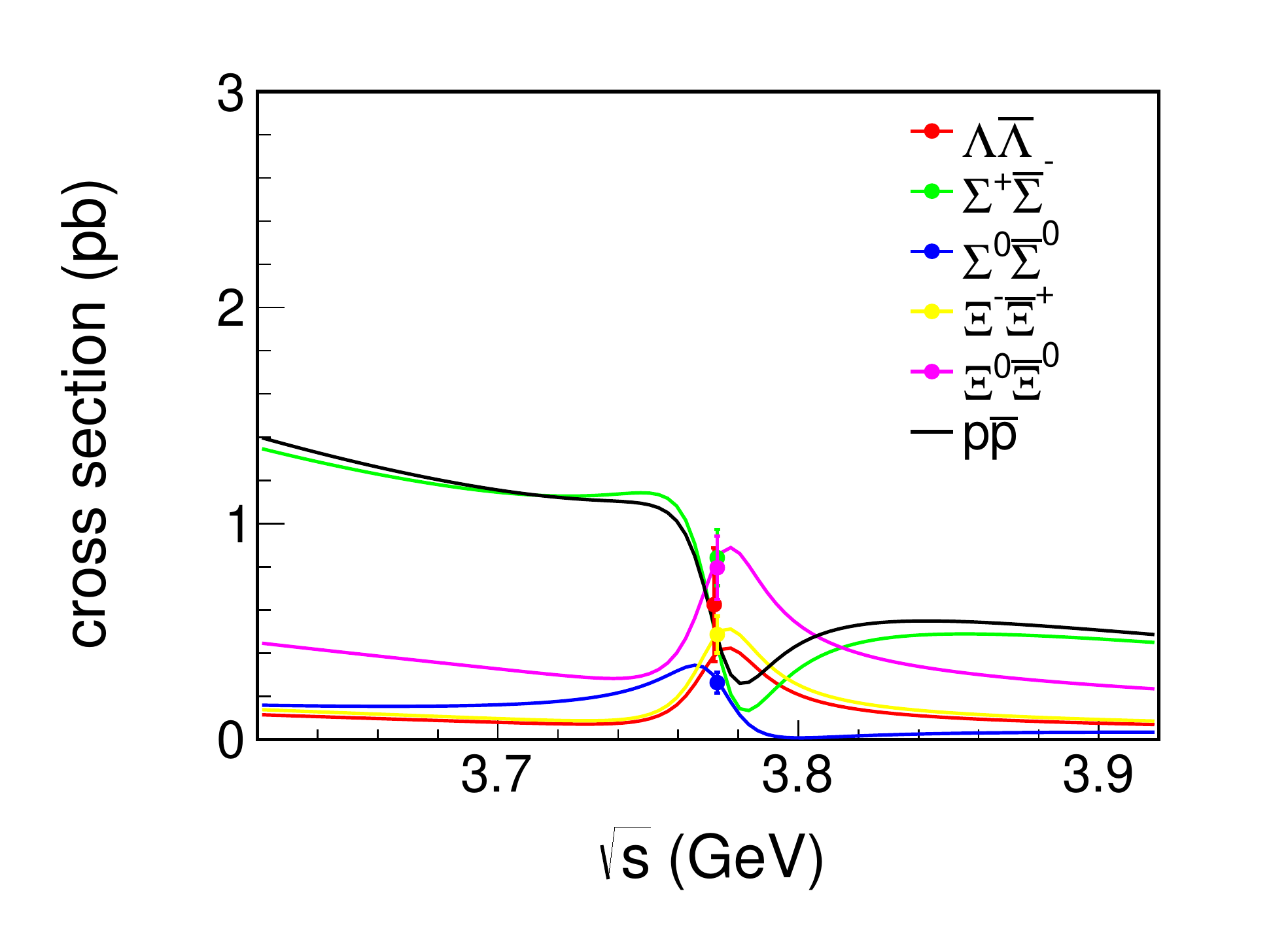}
   \put(-160, 130){\textbf{(b)}}\\
     \includegraphics[width = 0.45\textwidth]{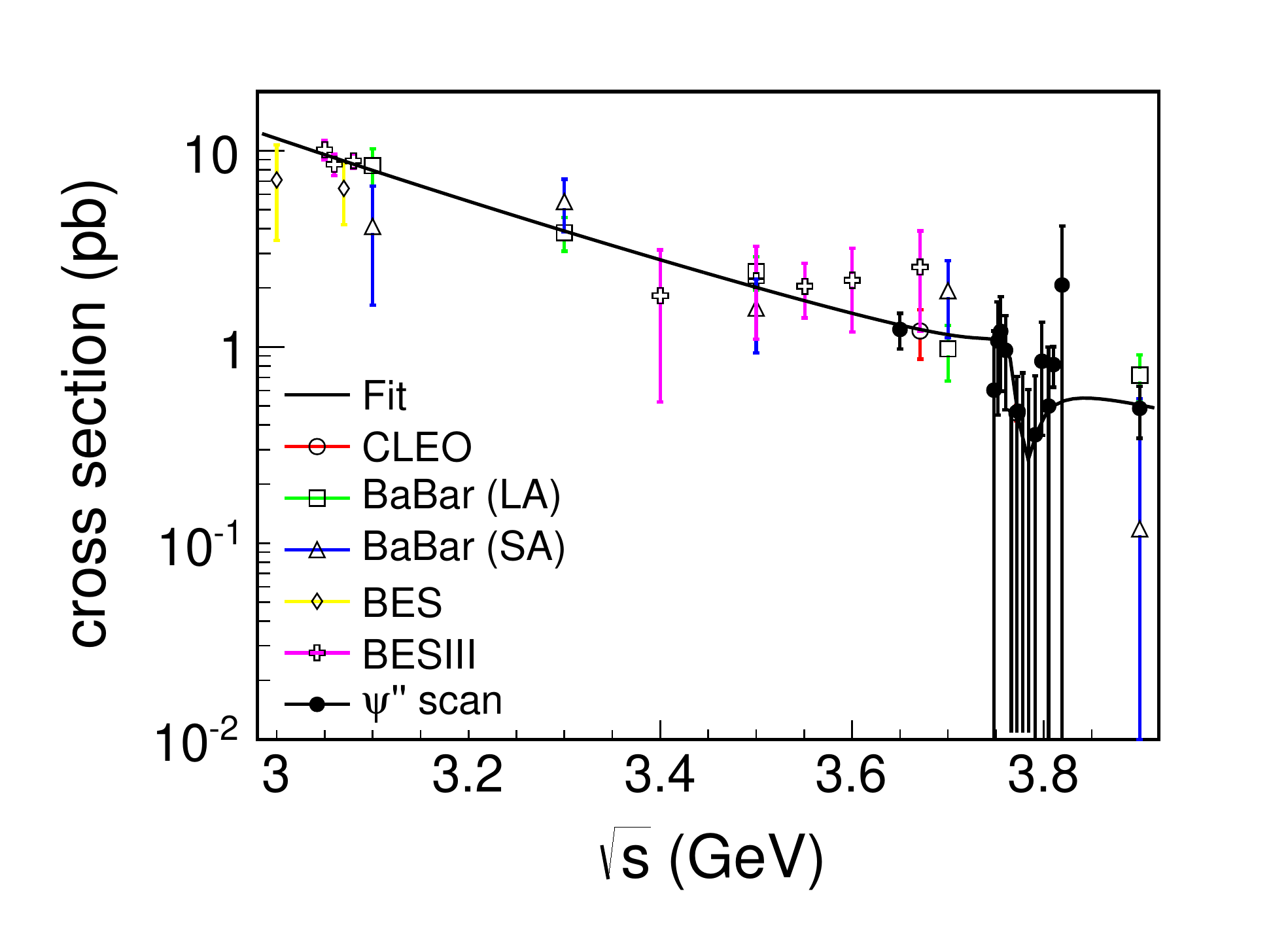}
   \put(-160, 130){\textbf{(c)}}
  \includegraphics[width = 0.45\textwidth]{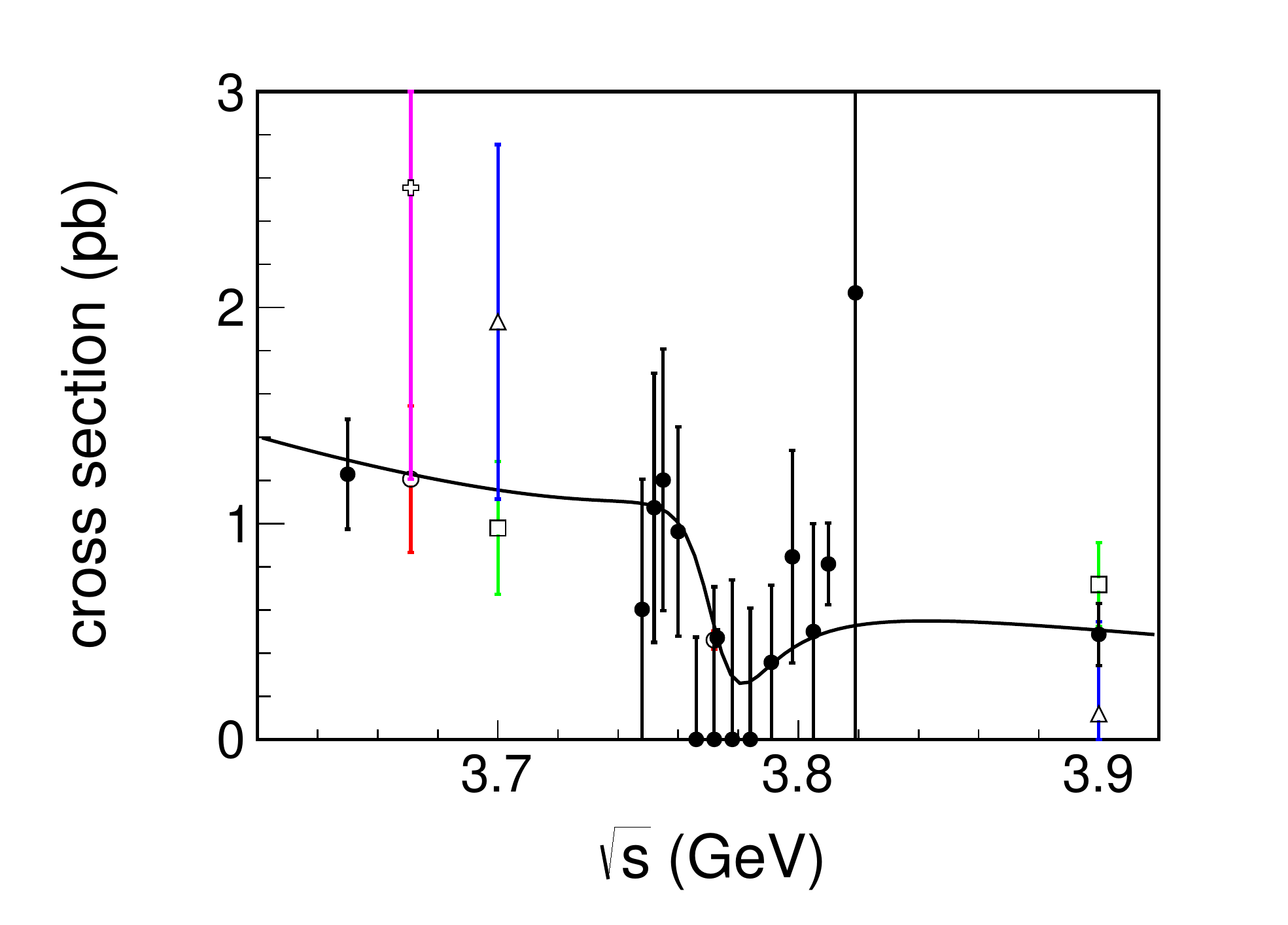}
   \put(-160, 130){\textbf{(d)}}\\
  \caption{The line shape of $\ep\en \to \BB$ and the fit. (b) and (d) are the zoom of (a) and (c) around the $\psi(3770)$ resonance, respectively. Dots with error bar are data. The curves are the theoretic prediction from the fit. In (a) and (b), the color-baryon correspondence relation is black-$p$, red-$\Lambda$, green-$\Sigma^+$, blue-$\Sigma^0$, yellow-$\Xi^-$ and pink-$\Xi^0$. The measurements of $\sigma(\ep\en\to\pp)$ are drawn in (c) and (d) while omitted in (a) and (b) for convenience.}\label{fig:fitBB}
\end{figure*}

Including all experiments, the fit results are summarized in Table~\ref{tab:fitBB1} and shown in Fig.~\ref{fig:fitBB}. The goodness of fit is $\chi^2/ndof = 43.7/31$. From Fig.~\ref{fig:fitBB}, we find that the line shape shows a dip structure around the $\psi(3770)$ resonance for the final states $\pp$ and $\Sigma^+\bar{\Sigma}^-$ and a bump structure for the final states $\Lambda\bar{\Lambda}$, $\Sigma^0\bar{\Sigma}^0$, $\Xi^-\bar{\Xi}^+$ and $\Xi^0\bar{\Xi}^0$.

\begin{table}[htbp]
\center{}
\caption{\label{tab:fitBB1} Fit results for $\ep\en \to \BB$}
\begin{tabular}{l l}
\hline\hline
Parameter & Value \\
\hline
$A_0$ & $2.91\pm0.22$ \\
$A_1$ & $0.37\pm0.15$ \\
$A_2$ & $0.07\pm0.30$ \\
$C_1$ (GeV$^4$) & $19.6 \pm 2.3$\\
$C_2$ (GeV$^4$) & $41.7 \pm 2.6$\\
$\phi'$ & $-54^\circ \pm 230^\circ$ \\
$\phi$ & $-137.5^\circ \pm 2.7^\circ$\\
\hline\hline
\end{tabular}
\end{table}

In Table~\ref{tab:fitBB1}, we find that $|A_1|, |A_2|<<|A_0|$, which means that the SU(3) breaking effect is small. In addition, the upper limits of $\sigma(\ep\en\to\Lambda\bar{\Lambda}/\Sigma^0\bar{\Sigma}^0/\Lambda\bar{\Sigma}^0)$ at $3.2-3.6$~GeV from the BABAR measurement~\cite{babarBB} are consistent with the predicted cross section from the fit result. Using the parameters from the fit, the branching fractions of $\psi(3770) \to \BB$ are calculated according to Eq.~\ref{eq:GammaB} and listed in Table~\ref{tab:fitBB2}. All $\BR(\psi(3770) \to \BB)$s are of the order of $10^{-5}$.

\begin{table}[htbp]
\center{}
\caption{\label{tab:fitBB2} Branching fraction of $\psi(3770) \to \BB$. The first uncertainty is from the global fit and the second uncertainty is due to the assumption that the electric and magnetic form factors are equal.}
\begin{tabular}{l c}
\hline\hline
Baryon  & $\BR(\psi(3770) \to \BB)$ ($\times10^{-5}$) \\
\hline
$\pp$ & $2.4\pm0.8\pm0.3$ \\
$\Lambda\bar{\Lambda}$ & $1.7\pm0.6\pm0.1$ \\
$\Sigma^+\bar{\Sigma}^-$ & $4.5\pm0.9\pm0.1$ \\
$\Sigma^0\bar{\Sigma}^0$& $4.5\pm0.9\pm0.1$ \\
$\Xi^-\bar{\Xi}^+$ & $2.0\pm0.7\pm0.1$ \\
$\Xi^0\bar{\Xi}^0$ & $2.0\pm0.7\pm0.1$ \\
\hline\hline
\end{tabular}
\end{table}

\subsection{Some discussions}
\begin{enumerate}
\item In the analysis above, the relaitons $|G_E|=|G_M|$ and $|F_E|=|F_M|$ are assumed. Fits are repeated assuming $|G_E|=0$ and $|F_E|=0$ instead, which is indicated from the measurement of the neutron form factors~\cite{fenice}. The branching fraction difference is taken as the systematic uncertainty (the second uncertainty term in Table~\ref{tab:fitBB2}).

\item We find that the two-solution ambiguity of $\BR(\psi(770) \to \pp)$ reported in Ref.~\cite{bes3psipp} is fixed with including the measurements about other baryon pairs. This can be clearly shown by comparing the $\chi^2$ curves as a function of the parameter $A^p$ using only the cross sections of $\ep\en \to \pp$ and using the cross sections of $\ep\en\to\BB$ ($B=p, \Lambda, \Sigma^+, \Sigma^0, \Xi^-$ and $\Xi^0$). The reduced $\chi^2$ curves are illustrated in Fig.~\ref{fig:Ap_scan}. The reduced $\chi^2$ is defined as the difference of the $\chi^2$ from the fit with $A^p$ fixed and that from the best fit. The blue curve in Fig.~\ref{fig:Ap_scan} indicates that the smaller solution in Ref.~\cite{bes3psipp} gives a better fit including all measurements of $\sigma(\ep\en\to\BB)$. In the case of assuming $|G_E|=0$ and $|F_E|=0$, this conclusion does not change.

\begin{figure}[htbp]
  \includegraphics[width = 0.45\textwidth]{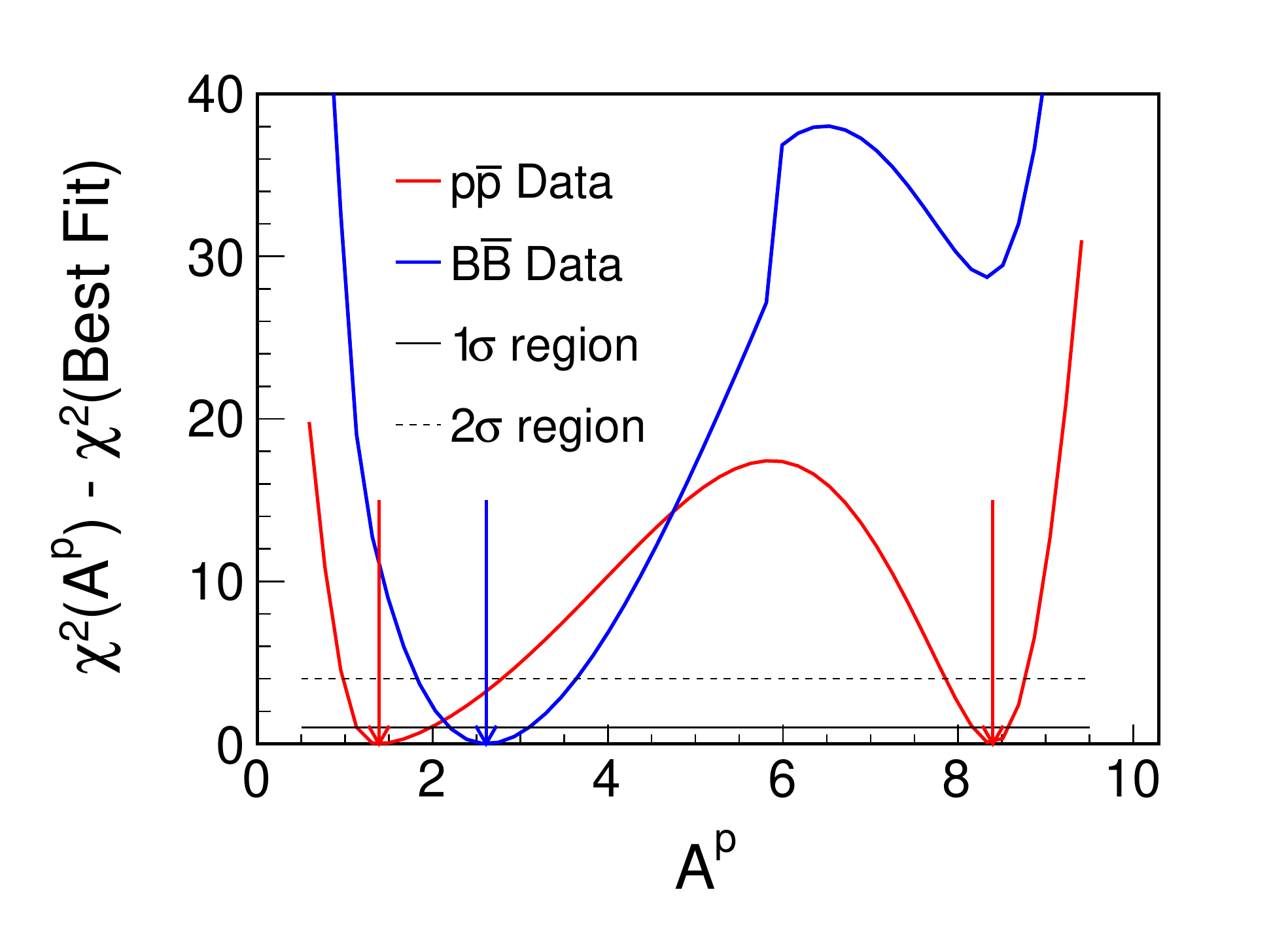}
  \caption{The curves of the reduced $\chi^2$ as a function of $A^p$ using only the cross sections of $\ep\en\to\pp$ (red curve) and using the cross sections of $\ep\en \to \BB$ (blue curve). The reduced $\chi^2$ is defined as the difference of the $\chi^2$ from the fit with $A^p$ fixed and that from the best fit. The arrows denote the best fits. The solid and dashed horizon lines denote the $1\sigma$ and $2\sigma$ regions respectively.}\label{fig:Ap_scan}
\end{figure}

\item The relative phase between the electromagnetic amplitude of the $\psi(3770)$ and the QED amplitude ($\phi'=-54^\circ \pm 230^\circ$) is consistent with 0 within the uncertainty. Fixing $\phi'=0$ produces negligible effect. The relative phase of the OZI-suppressed strong decay amplitude of the $\psi(3770)$ and the QED amplitude is found to be $-137.5^\circ \pm 2.7^\circ$. Many phenominological analyses~\cite{kzhu, phase0, phase1, phase2, phase3, phase4, phase5, phase6, phase7} have been performed for various final states in the hadronic decays of $\jpsi$ and $\psi(3686)$. It is revealed that the relative phase $\phi$ is close to $-90^\circ$. For $\jpsi/\psi(3686) \to \BB$, two possible phase values are found using a similar model in Ref.~\cite{kzhu}. If the relative phase is assumed to be universal whatever the final state is, the large negative values are favored and close to $-90^\circ$ for $\jpsi$ and $\psi(3686)$. 
For the decay mode $\psi(3770) \to p\bar{p}$, the calculation of Ref.~\cite{phase_pp} shows that the dominant contribution is the OZI-suppressed amplitude and $\phi=-113^\circ$. If the contribution of the OZI-allowed $D\bar{D}$ state as an intermediate state (which is firstly introduced in Ref.~\cite{lipkin}) is included, the phase angle $\phi$ becomes $-99^\circ$. 
However, our study shows the phase is far from $-90^\circ$, which indicates that there may be additional mechanism contributing to the baryon-pair decays of $\psi(3770)$. 

\begin{table}[htbp]
\center{}
\caption{\label{tab:phase} The relative phase between the amplitude of the strong interaction and that of the electromagnetic interaction in the decays of $\psi \to \BB$.}
\begin{tabular}{l c c}
\hline\hline
Charmonium & $\phi'$ & $\phi$\\
\hline
$\jpsi$ & fixed at 0  & $-85.9^\circ\pm1.7^\circ$ or $+90.8^\circ\pm1.6^\circ$~\cite{kzhu}\\
$\psi(3686)$ & fixed at 0 & $-98^\circ\pm25^\circ$ or $+134^\circ\pm25^\circ$~\cite{kzhu}\\
$\psi(3770)$ & $-54^\circ \pm 230^\circ$ & $-137.5^\circ\pm2.7^\circ$ \\
\hline\hline
\end{tabular}
\end{table}

\item Assuming the probability of a charmonium state $\psi$ decaying to light hadrons is proportional to the absolute square of the value of the charmonium wave function at the origin, we can relate the branching fractions of $\jpsi/\psi(3686)/\psi(3770) \to \BB$. We define a dimensionless quantity, $\kappa_\psi^B$, in Eq.~\ref{eq:kappa}.
\begin{equation}\label{eq:kappa}
  \kappa_\psi^B \equiv \frac{\Gamma(\psi\to\BB)/\beta_{\psi}^B}{\Gamma_e(\psi)} = \frac{\BR(\psi\to\BB)/\beta_{\psi}^B}{\BR_e(\psi)}\: ,
\end{equation}
where $\beta_{\psi}^B=\sqrt{1-4M_B^2/M_\psi^2}$ and $\Gamma_e(\psi)$ ($\BR_e(\psi)$) is the partial width (branching fraction) of $\psi\to\ep\en$. Under the assumption above, it is expected that $\kappa_{\jpsi}^B = \kappa_{\psi(3686)}^B=\kappa_{\psi(3770)}^B$ and that
\begin{eqnarray}
&&\BR(\psi(3770) \to \BB) = \frac{\kappa_{\psi(3770)}^B}{\kappa_{\psi}^B}\BR_{sc.}(\psi(3770)\to \BB) \label{eq:BRpsipp}\\
&&\BR_{sc.}(\psi(3770)\to \BB)\equiv\BR(\psi\to\BB)\frac{\BR_e(\psi(3770))}{\BR_e(\psi)}\frac{\beta_{\psi(3770)}^B}{\beta_{\psi}^B} \:,  \nonumber
\end{eqnarray}
where $\psi =\jpsi$ or $\psi(3686)$, and $\BR_{sc.}(\psi(3770) \to \BB)$ is from a scaling of $\BR(\psi\to\BB)$ under the assumption $\kappa_{\psi(3770)}=\kappa_{\psi}^B$.

Table~\ref{tab:kappa} lists the $\kappa_\psi^B$s for $\psi = \jpsi, \psi(3686)$ and $\psi(3770)$. We find that $\kappa_{\jpsi}^B \simeq \kappa_{\psi(3686)}^B < 0.1 \kappa_{\psi(3770)}^B$, which means that $\BR(\psi(3770)\to\BB)$ is at least one order of magnitude larger than that scaled from $\BR(\jpsi/\psi(3686) \to \BB)$, as shown in Eq.~\ref{eq:BRpsipp}. 
%It may shed light on the puzzle about the non-$D\bar{D}$ branching fraction of $\psi(3770)$.

\begin{table}[htbp]
\center{}
\caption{\label{tab:kappa} $\kappa$ values for $\psi\to \BB$. For $\jpsi$ and $\psi(3686)$, the experimental measurements included in the PDG~\cite{PDG}  are used.}
\begin{tabular}{c l l l}
\hline\hline
Baryon pair & $\kappa_{\jpsi}$ & $\kappa_{\psi(3686)}$ & $\kappa_{\psi(3770)}$\\
\hline
$\pp$ & $0.045\pm0.001$ & $0.041\pm0.002$ & $2.9\pm0.9$\\
$\Lambda\bar{\Lambda}$ & $0.039\pm0.004$ & $0.045\pm0.008$ & $2.2\pm0.8$\\
$\Sigma^+\bar{\Sigma}^-$ & $0.039\pm0.006$ & $0.043\pm0.013$ & $6.1\pm1.3$\\
$\Sigma^0\bar{\Sigma}^0$& $0.034\pm0.002$ & $0.037\pm0.007$ & $6.1\pm1.3$\\
$\Xi^-\bar{\Xi}^+$ & $0.027\pm0.004$ & $0.033\pm0.011$ & $2.9\pm1.0$\\
$\Xi^0\bar{\Xi}^0$ & $0.024\pm0.003$ & $0.051\pm0.016$ & $3.0\pm1.0$\\
\hline\hline
\end{tabular}
\end{table}

\item In view of last point, it is necessary to have a small review about the non-$D\bar{D}$ decay modes which have been observed experimentally. Table~\ref{tab:review} summarizes the measured partial width of these non-$D\bar{D}$ decay modes and the theoretical predictions. The measured partial width is calculated by multiplying the full decay width by the corresponding measured branching fraction~\cite{PDG}. From Table~\ref{tab:review}, we find that the potential models proposed in Ref.~\cite{jpsipipi_kuang} and Ref.~\cite{gammachicj_ding, gammachicj_eithen, gammachicj_barnes} can explain well the rate of the decay modes with the charmonium final state. These models assume that $\psi(3770)$ and $\psi(3686)$ are the mixture of the $2S$ and $1D$ states of the $c\bar{c}$ system, as shown in Eq.~\ref{eq:mix} with the mixing angle $\theta \simeq -10^\circ$.  
\begin{eqnarray}
\psi(3686) &=& \psi_{2S}\cos\theta + \psi_{1D}\sin\theta \nonumber \\
\psi(3770) &=&-\psi_{2S}\sin\theta + \psi_{1D}\cos\theta \label{eq:mix}
\end{eqnarray}
\begin{table}[htbp]
\center{}
\caption{\label{tab:review} Partial width of the observed non-$D\bar{D}$ decay modes of the $\psi(3770)$. $\Gamma_{meas.}$ ($\Gamma_{pred.}$) represents the measured (predicted) partial width.}
\begin{tabular}{l l l}
\hline\hline
Decay Mode & $\Gamma_{meas.}$ (keV) & $\Gamma_{pred.}$ (keV) \\
\hline
$\jpsi\pip\pim$ & $52.5\pm7.6$ & 20-110~\cite{jpsipipi_kuang}\\
$\jpsi\piz\piz$ & $21.8\pm8.2$ & \\
$\jpsi\eta$ & $24\pm 11$& \\
%$\gamma\chi_{c0}$ & $198.6\pm24.5$& 199~\cite{gammachicj_ding},524~\cite{gammachicj_rosner},225~\cite{gammachicj_eithen},213~\cite{gammachicj_barnes}\\
%$\gamma\chi_{c1}$ & $73.4\pm13.6$ & 72~\cite{gammachicj_ding},73~\cite{gammachicj_rosner},59~\cite{gammachicj_eithen},77~\cite{gammachicj_barnes}\\
$\gamma\chi_{c0}$ & $198.6\pm24.5$& 199-225~\cite{gammachicj_ding, gammachicj_eithen, gammachicj_barnes}, 524~\cite{gammachicj_rosner}\\
$\gamma\chi_{c1}$ & $73.4\pm13.6$ & 59-77~\cite{gammachicj_ding, gammachicj_rosner,gammachicj_eithen,gammachicj_barnes}\\
$e^+e^-$ & $0.262\pm0.018$ & \\
$\phi\eta$ & $8.4\pm1.9$ & \\
\hline\hline\end{tabular}\end{table}

However, for exclusive light hadron decay modes, it is difficult to have an accurate theoretical prediction. In this work, the combined branching fraction of $\psi(3770) \to B\bar{B}$ is of the order of $10^{-4}$ by summing up the numbers in Table~\ref{tab:fitBB2}. Though it is much smaller than the non-$D\bar{D}$ branching fraction of the order of about 10\% measured by the BES collaboration~\cite{nonDDbes1,nonDDbes2,nonDDbes3}, the baryon pairs only account for a small fraction of the light hadron decay modes. Furthermore, this work shows $\BR(\psi(3770)\to\BB)$ is at least one order of magnitude larger than that scaled from $\BR(\jpsi/\psi(3686) \to \BB)$ as discussed above. This indicates that the mechanism in the light-hadron decays of the $\psi(3770)$ is different from that in the case of the $\jpsi$ or $\psi(3686)$. 

\end{enumerate}

\section{Summary}
Focusing on one type of non-$D\bar{D}$ decays, $\psi(3770)$ into baryon anti-baryon pair, all available experiments of measuring the cross section of $\ep\en\to\BB$ at c.m. energy ranging from 3.0~GeV to 3.9~GeV are collected. A model based on the SU(3) flavor symmetry is built to relate the baryon octets. The SU(3) breaking effects due to the electromagnetic interaction and the quark mass difference are also considered. A global fit based on this model is performed. The two-solution ambiguity about $\BR(\psi(3770) \to \pp)$ reported in Ref.~\cite{bes3psipp} is fixed. We find that the statistical significance of the presence of the process $\ep\en\to\psi(3770) \to \pp$ is much larger than 5 standard deviations, which is not reported in Ref.~\cite{bes3psipp}. $\BR(\psipp \to \BB)$ is determined to be $\BRp$, $\BRlam$, $\BRsigp$, $\BRsigz$, $\BRxim$, and $\BRxiz$ for $B=p, \Lambda, \Sigma^+, \Sigma^0, \Xi^-$ and $\Xi^0$, respectively. They are at least one order of magnitude larger than a simple scaling of $\BR(\jpsi/\psi(3686) \to \BB)$. Furthermore, the relative phase between the strong ampltitude and the electromagnetic amplitude is found be to far from $-90^\circ$, which are favored in the hadronic decays of $\jpsi$ and $\psi(3686)$. The two evidences above may shed light on the puzzle about the non-$D\bar{D}$ branching fraction of $\psi(3770)$.

\section{Acknowledgements}
Li-Gang Xia would like to thank Fang Dai for many helpful discussions.

\end{document}